\documentclass[preprint,authoryear,12pt]{elsarticle}
\usepackage{lineno,hyperref,graphicx}

\usepackage{amssymb}

\journal{Advances in Space Research}

\begin{document}

\begin{frontmatter}



\title{Space VLBI: from first ideas to operational missions}


\author{Leonid I. Gurvits}
\address{Joint Institute for VLBI ERIC, 
               \\ Oude Hoogeveensedijk 4, 7991 PD Dwingeloo, The Netherlands \\
                and \\
                Department of Astrodynamics and Space Missions, Delft University of Technology, \\
                Kluyverweg 1, 2629 HS Delft, The Netherlands  }
\ead{lgurvits@jive.eu}


\begin{abstract}

The operational period of the first generation of dedicated Space VLBI (SVLBI) missions commenced in 1997 with the launch of the Japan-led mission VSOP/HALCA and is coming to closure in 2019 with the completion of in-flight operations of the Russia-led mission RadioAstron. They were preceded by the SVLBI demonstration experiment with the Tracking and Data Relay Satellite System (TDRSS) in 1986--1988. While the comprehensive lessons learned from the first demonstration experiment and two dedicated SVLBI missions are still awaiting thorough attention, several preliminary conclusions can be made. This paper addresses some issues of implementation of these missions as they progressed over four decades from the original SVLBI concepts to the operational status.

\end{abstract}


\end{frontmatter}

\parindent=0.5 cm

\section{Introduction}

For more than half a century, since its first demonstrations in 1967, Very Long Baseline Interferometry (VLBI) technique holds the record in sharpness of studying astronomical phenomena. Paradoxically, this record is achieved in the long-wavelength domain of electro-magnetic spectrum (comparing to other bands of spectrum used in astronomy studies -- infrared, optical, ultraviolet, X- and gamma-ray). This goes against the fundamental physical dependence defined by diffraction -- the longer the wavelength, the lower the angular resolution for a fixed size of the receiving aperture. The reason of this apparent inconsistency is in the technological difficulties of reaching the diffraction limit at short wavelengths and high quantum energies. In contrast, in radio domain, diffraction limit can be reached by relatively simple technical means.

Radio astronomy, born in 1933, is one of the pillars of the modern ``all wavelength'' astrophysics. During the first two decades of its development, while offering a new and at times key vantage point for understanding astrophysical phenomena, radio astronomy was behind its much older optical counterpart in terms of angular resolution (``sharpness of vision'') due to the diffraction limit of ``single dish'' antennas. For centimeter to meter wavelengths, affordable dishes of tens to a hundred of meters in diameter could reach angular resolution of tens to single-digit minutes of arc, far worse than a typical arcsecond angular resolution of Earth-based optical telescopes (limited by the Earth's atmosphere, not even the instrumental diffraction). A concept of interferometers was ``imported'' into radio astronomy from optics in the beginning of the 1950s \citep[e.g.][Ch. 5]{Bur-Smi}. It enabled radio astronomers to sharpen the angular resolution proportionally to the ratio of the interferometer baseline to the single dish diameter. Later, in the second half of the 1960s, the radio interferometers got their ultimate extension to the baselines, comparable to the Earth diameter. This was the beginning of VLBI \citep{Moran98}, which enabled radio astronomers to reach record-high angular resolution of milliarcseconds and even sub-milliarcseconds. The record in angular resolution is held firmly by VLBI not only among all astronomical techniques in all domains of electromagnetic spectrum but arguably among all branches of experimental science. However, the limit of VLBI angular resolution defined by the Earth diameter was not unnoticed. In fact, even before getting first VLBI fringes in ``live'' observations in 1967, a manuscript of the paper by \citet{MaKaSh65} prepared in 1963, in which the idea of VLBI was presented for the first time, had a paragraph suggesting a radio telescope on a spacecraft enabling baselines even longer than the Earth diameter (G.B.~Sholomitsky, 1987; N.S.~Kardsashev, 1994; L.I.~Matveenko, 2012; private communications). Not surprisingly, soon after the demonstration of first VLBI fringes, a push for baselines longer than the Earth diameter materialised in a number of design studies of Space VLBI (SVLBI) systems.

Over the past half a century, several dozens of various SVLBI concepts have been presented with widely varying depth of development and level of detalisation. Several of them reached a pre-design study stage in the 1970s -- early 1990s. Among them, the most prominent were the European-led concept QUASAT \citep{RTS84}, the Soviet project KRT-30 as a part of the RACSAS system (\citet{NSK+80}, \citet{RZS84},  \citet[Ch. 9]{Davi97} and \citet[Ch. 8]{HenVis}) and the International VLBI Satellite (IVS) project \citep{Pilb91}. These studies were not selected by respective space agencies for implementation, and therefore were not subjected to full engineering analysis and development. However, they paved the way for the first SVLBI demonstration in the middle of 1980s (TDRSS Orbital VLBI experiment, \cite{Levy+86}) and two dedicated SVLBI missions, VSOP/HALCA launched in 1997 \citep{Hir98} and RadioAstron launched in 2011 \citep{NSK+13}.

\begin{figure}[t]
   \centering
\includegraphics[width=62mm,angle=90]{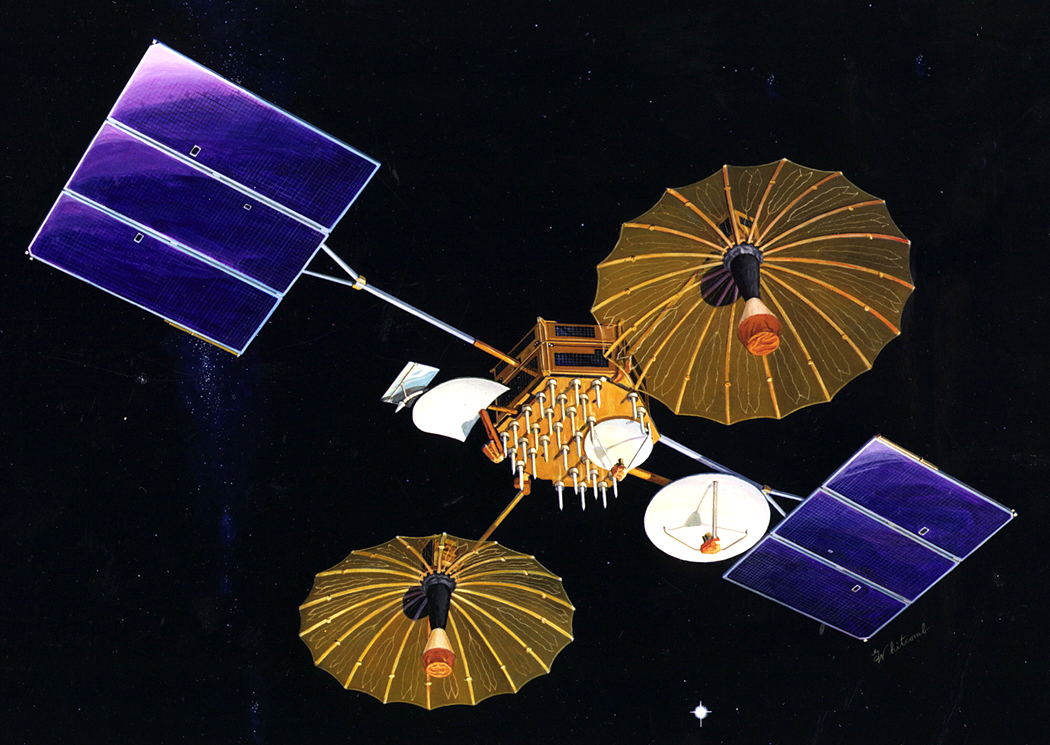}
\includegraphics[width=62mm,angle=90]{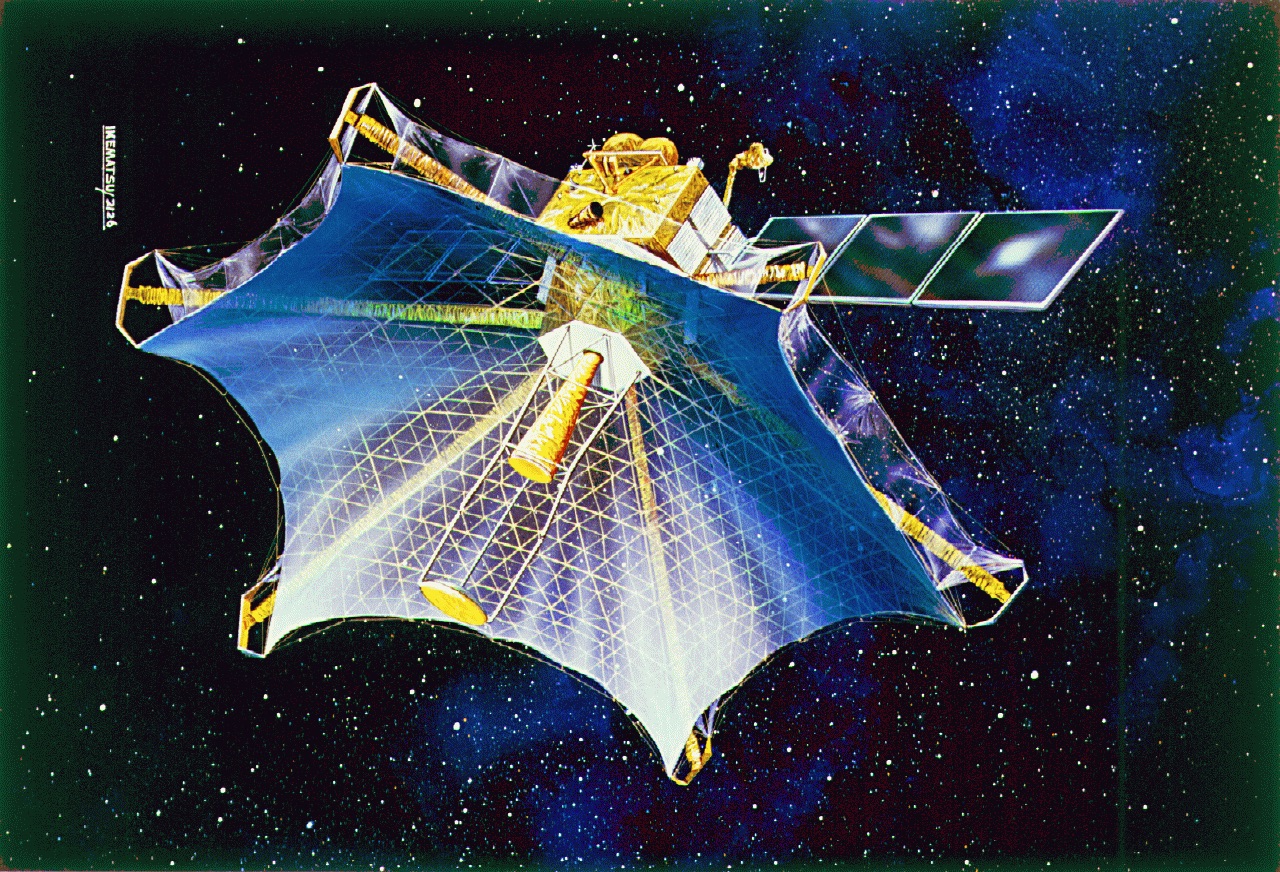}
\includegraphics[width=62mm,angle=90]{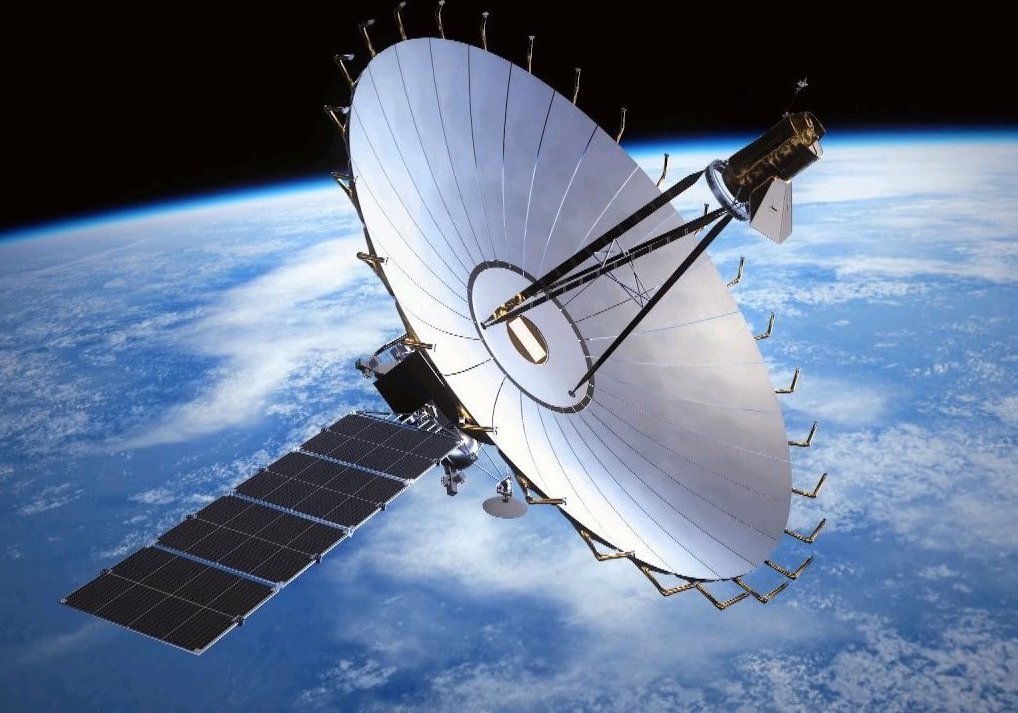}
\caption{Artist's impressions of: ({\sl left panel}) Tracking and Data Relay Satellite System (TDRSS) spacecraft, picture credit: NASA; ({\sl central panel}) Highly Advanced Laboratory for Communication and Astronomy (HALCA) of the VLBI Space Observatory Program (VSOP), picture credit: JAXA;  ({\sl right panel}) Spektr-R spacecraft of the RadioAstron mission, picture credit: Lavochkin Scientific and Production Association.}
   \label{fig-3_SVLBI}   
\end{figure}

First VLBI ``fringes'' on baselines longer than the Earth diameter were obtained with the NASA’s geostationary spacecraft of the Tracking and Data Relay Satellite System (TDRSS, shown in the left panel of Fig. \ref{fig-3_SVLBI}) in 1986 \citep{Levy+86}. This was a very efficient example of ad hoc use of existing orbiting hardware not designed originally for conducting SVLBI observations. The main outcome of several TDRSS observing campaigns was two-fold. First, the very concept of getting coherent interferometric response (the so called interferometric ``fringes'') on baselines longer than the Earth diameter has been proved experimentally. Second, observations of two dozens of strongest AGN (Active Galactic Nuclei) at 2.3 GHz in 1986--87 \citep{Levy+89,Linf+89}, and in a dual-frequency mode at 2.3 and 15 GHz in 1988 \citep{Linf+90} provided indications that at least some extragalactic sources of continuum radio emission were more compact and therefore brighter than expected. These milestones supported growing momentum for the first generation of dedicated Space VLBI missions. 

At the time of this writing, the first generation SVLBI era has practically come to the completion. The VSOP/HALCA operated in orbit in the period 1997--2003. Its science heritage is summarised in \citet{Hirax+2000b}, \citet{Hirax+2000c}, \citet[Parts 3-4]{Hagi+09}. The RadioAstron mission was operational in the period 2011--2019. Its science outcome is still to be worked out; some preliminary summaries are presented in \citet{NSK+YYK17} and in the present Special Issue. A very brief and subjective list of major legacy achievements of VSOP and RadioAstron missions with associated references is given in \citet{LIG-18}. The present paper does not set out the science results of VSOP and RadioAstron which can be found in the mentioned references; rather, it presents a brief attempt to summarise experience and lessons learned from development and in-flight operations of the first generation Space VLBI missions.

\section{Science case of the first generation SVLBI}

At the pre-design phase, both VSOP and RadioAstron projects formulated a range of science tasks, all driven by the need for improving the angular resolution limited by the Earth diameter. Arguably, the strongest case dealt with the necessity of experimental verification of the highest brightness of synchrotron emission in AGN -- the upper limit of brightness temperature $T_B \approx 10^{11.5}$~K defined by the Inverse Compton Scattering \citep{KelPau69} and a stricter upper limit $T_B \approx 10^{10.5}$~K defined by the energy equipartition between electrons and magnetic field \citep{Read94}. These two limits are consistent with the canonical model of AGN radio emission; if they don't hold, the model is in great doubts. It is a rather curious case of a ``cosmic conspiracy'': our planet is ``tuned'' in a very special way for tackling the AGN brightness temperature quest. Brightness temperature of a source of the flux density $S$ at redshift $z$ slightly resolved with a synthesised interferometric beam of elliptical shape with HPBW major and minor axis $\theta_{\rm maj}$ and $\theta_{\rm min}$, respectively, is given by the following approximation:

\begin{equation} \label{t-bright}
T_{\rm B}=\frac{2 \ln{2}}{\pi k}S(1+z)\frac{\lambda^{2}}{\theta_{\rm maj}\theta_{\rm min}} \propto \frac{2 \ln{2}}{\pi k}S(1+z)B^{2} \,\,\,\, ,
\end{equation}

\noindent where $k$ is Boltzmann constant and $B$ is a characteristic baseline of an interferometer. For the longest possible baseline on Earth of about $10^{4}$\,km and a source of a flux density 500\,mJy at $z=1$, the highest brightness temperature that can be measured directly, not as a lower limit, as defined by equation (\ref{t-bright}), is about $10^{12}$\,K, exactly coinciding with the theoretical Inverse Compton limit. Thus, in order to verify the Inverse Compton limit experimentally one has to operate with baselines longer than the Earth diameter.

The first generation SVLBI projects aimed at more than verification of brightness temperature limits of synchrotron emission in AGN. These included spectral line studies of masers at 1.35 and 18~cm, investigations of pulsar emission and impacts of the interstellar medium and interstellar scintillation, as well as other topics. An overview of the VSOP science case was presented at the International Symposium held at the Institute of Astronautical and Space Science (Sagamihara, Japan) in December 1989 \citep{Hirax+91}. The science case of the RadioAstron evolved over the long pre-launch history. Its most comprehensive presentation was published two years prior to the launch \citep{NSK-09}.

\section{Hardware bottlenecks of SVLBI}

Space VLBI is no different from ``traditional'' Earth-based VLBI systems in terms of principle composition of instrumentation at each telescope. In the case of a space-borne radio telescope, its instrumentation must be space-qualified. It drives up the cost of the telescope hardware. However, it does not pose unsurmountable problems -- provided funding for its development, tests and construction is sufficient. Nevertheless, three subsystems of SVLBI are more challenging than others. They define the overall Technology Readiness Level (TRL) of a concept, and ultimately -- the fate of the mission at the selection stage as well as its scientific value. These three subsystems are the main radio telescope antenna, local heterodynes and VLBI data transport instrumentation.  

\begin{figure}[t]
   \centering
\includegraphics[width=92mm,angle=0]{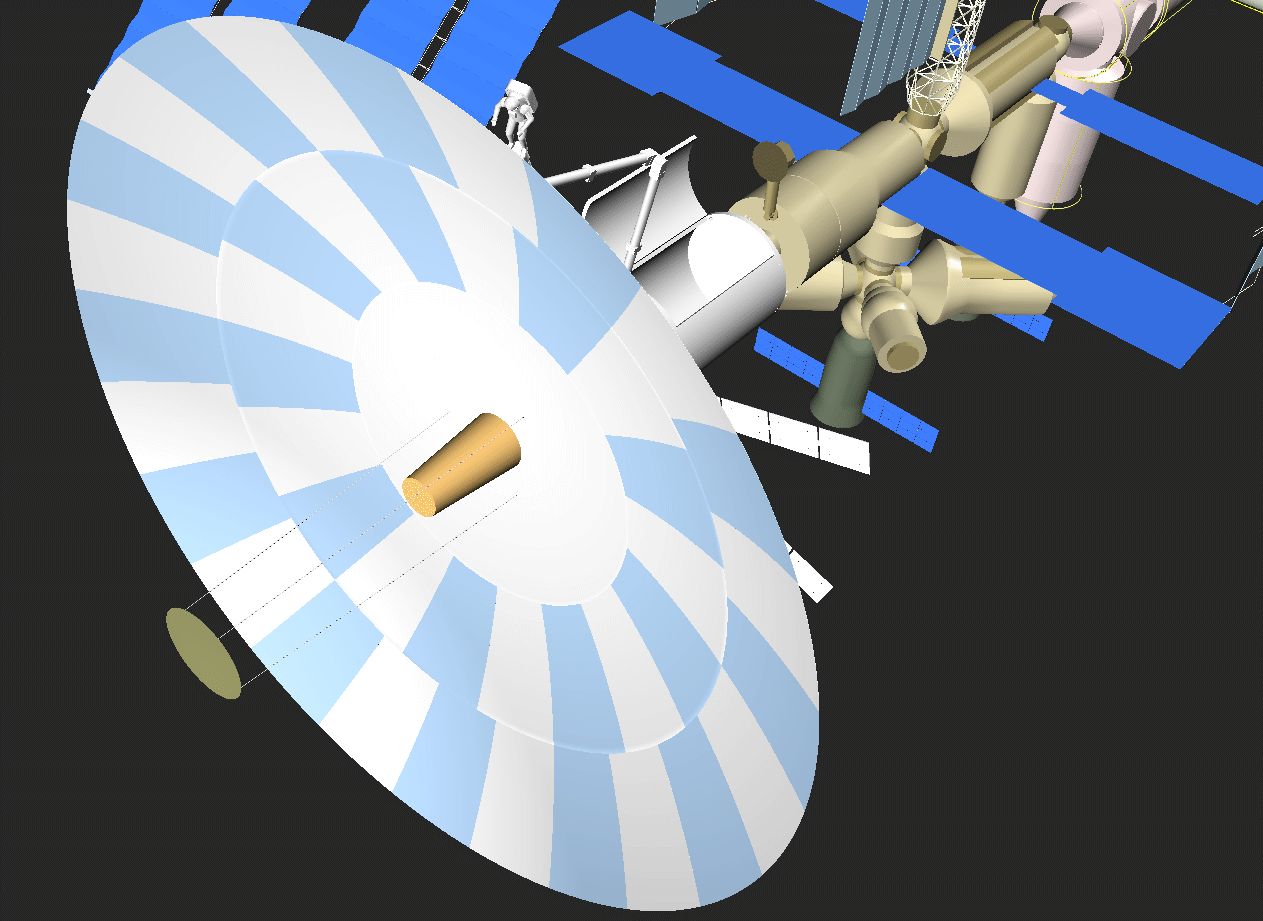}
\caption{An artist's impression of a large radio telescope (diameter 30~m or larger) assembled at a near-Earth orbital station like the International Space Station (ISS). After completion of assembly and tests, the telescope de-docks from the station and drifts to an operational SVLBI orbit, possibly using low-thrust jets to lower the acceleration load on a large deployed antenna reflector. Image courtesy Aerospatiale, 1998.}
   \label{fig-Aerospa}   
\end{figure}

\subsection{Space-borne VLBI antennas}  \label{SVLBI-ant}

The major critical component of a space-borne VLBI observatory is the space-borne radio telescope antenna -- its mechanical and electrical subsystems. For both first generation SVLBI missions VSOP/HALCA and RadioAstron, the diameter of the main antenna reflector was essentially defined by the space launcher capability. In turn, the antenna diameter, about 10~m, was the main factor in defining the overall sensitivity of the SVLBI mission. In all VSOP and RadioAstron observations, the space-borne antenna was the smallest in an array, at least 2.5 times smaller in diameter than the smallest co-observing Earth-based antenna. Not surprisingly, SVLBI observations with VSOP and RadioAstron were much less sensitive than contemporary Earth-based VLBI experiments. Larger apertures of space-borne elements of SVLBI systems have been considered in the 1980s-90s: QUASAT, 15\,m, \citep{RTS84}; IVS, 25\,m \citep{Pilb91}; ARISE, 25--30\,m \citep{UlvGuLi97,Ulve99}; KRT-30, 30\,m \citep{NSK+80}. In the two implemented SVLBI missions as well as in all mentioned concept studies, the main reflector of the telescope was ``pushing'' the launch capability to the limit. It is rather realistic to expect that new launchers would be able to place on Earth orbits 30-m-class reflectors in the next several decades. However, a true breakthrough in constructing large apertures of space-borne radio telescopes would come with assembling them in orbit. Such the assembly might be conducted while a radio astronomy spacecraft is docked temporarily to an orbital station \citep{LIG2000}, Fig.~\ref{fig-Aerospa}. An ultimate solution for an unlimited size of the space-borne radio telescope has been suggested four decades ago, an infinitely expandable aperture in space \citep{Buyakas79} which probably can be referred to in the modern terminology as a  ``Square Kilometre Array in Space''. 

\subsection{Local VLBI heterodynes}

Sufficiently stable heterodyning is an indispensable attribute of a VLBI system. In ``traditional'' VLBI, heterodynes are fed by local highly stable oscillators (frequency standards) colocated with telescopes. A rule of thumb for their stability requirement is given by the following formula \citep{RogMor81}:

\begin{equation} \label{sig-tau}
\nu_{\rm max} \approx  [\tau \sigma_{y}(\tau)]^{-1} \,\,\,\, ,
\end{equation}

\noindent where $\nu_{\rm max}$ is the maximum observing frequency possible without significant loss of coherency, $\tau$ is integration time and $\sigma^{2}_{y}(\tau)$ is the two-sample Allan variance representing the instability of the local oscillator (LO). For a realistic integration time of 10--100~s (essentially determined by scattering for Earth-based telescopes) at cm wavelengths, the required stability of the local oscillator defined by the equation (\ref{sig-tau}) is $\sigma_{y}(\tau) = 3\times 10^{-13}$. This value is well within specifications of laboratory hydrogen maser (H-maser) oscillators; most modern Earth-based VLBI telescopes are equipped with H-masers. 

\begin{figure}[t]
   \centering
\includegraphics[width=82mm,angle=0]{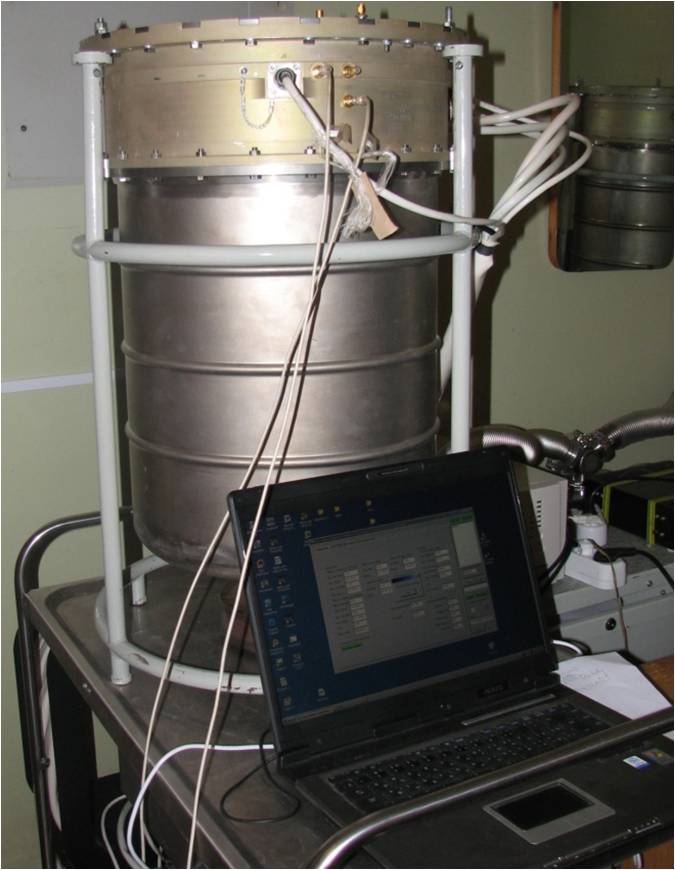}
\caption{One of the RadioAstron's on-board H-maser units undergoing laboratory tests, Moscow, Astro Space Center of P.N. Lebedev Physical Institute, July 2008.}
   \label{H-maser}   
\end{figure}

In SVLBI, a stable heterodyning can be provided to a space-borne radio telescope either by an on-board stable oscillator or by transferring a sufficiently stable signal from the Earth-based station via a phase-locked loop (PLL). The former approach obviously requires a space-qualified local oscillator device. The RadioAstron spacecraft Spektr-R is equipped with two on-board H-maser units (Fig. \ref{H-maser}), which are the first active H-maser oscillators flown on a space science mission. Cheaper and less demanding rubidium frequency standards, although inferior to H-masers in terms of stability at typical for VLBI integration times, can be used for lower frequency VLBI experiments, and RadioAstron has one of these too as a part of its science payload.

The PLL option was investigated in details by  \citet{VVAndr82} and \citet{LRD91}. It was implemented as a secondary LO system in the RadioAstron project. VSOP operated with PLL only, as the HALCA spacecraft was not equipped with a VLBI-qualified on-board LO. 

A convenient measure of coherency in VLBI is a dependence of a signal-to-noise ratio (SNR) of fringe detection on the integration time (\citet{TMS2017}, section 9.5.3, Fig.~9.16). In Earth-based VLBI systems at cm-dm wavelengths, modern H-masers do not limit increase of fringe SNR with increasing integration time to values from $\sim3\times 10^{3}$ to $\sim3\times 10^{4}$~s. In practice, the integration time is reduced by propagation effects in ionosphere (at frequencies lower than $\sim$10~GHz) and atmosphere (at frequencies higher than $\sim$20~GHz). As demonstrated by VSOP, the PLL synchronisation enabled observations with coherent integration of the order of 300~s \citep{Hirax+2000a}. In RadioAstron observations, the same coherent integration has been achieved at all four observing bands, both using the on-board H-maser and PLL synchronisation (\cite{Likh+2017} and Y.Y.~Kovalev, private communication, 2018). It has to be noted that the PLL synchronisation option requires continuing two-way radio link with spacecraft. Also, it requires accounting for instrumental delay introduced by the PLL link while cross-correlating VLBI data streams from the space-borne and Earth-based telescopes. As demonstrated by both VSOP and RadioAstron, these complications can be addressed successfully. Thus, one could conclude that inclusion in the payload of a SVLBI spacecraft operating at centimetre and longer wavelengths such a complicated and expensive device as a space-qualified H-maser is unnecessary provided the PLL option can be used. That said, the presence of the H-maser on board the RadioAstron spacecraft made it possible to conduct the ad hoc Gravitational Redshift Experiment \citep{Litv+18,Nunes+19}. Future advanced SVLBI systems operating at millimetre and sub-millimetre wavelengths  \citep{Andria+19,Fish+19,Kudria+19,Linz+19,Roelofs+19} might require new approaches to space-borne telescope heterodyning. 

\begin{figure}[t]
   \centering
\includegraphics[width=142mm,angle=0]{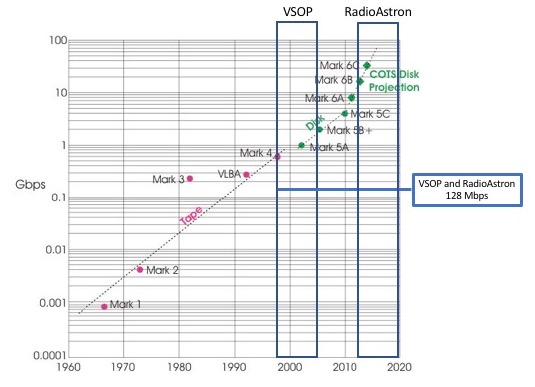}
\caption{The rate of VLBI data streams per telescope as a function of time over the 50-year history of VLBI. The points and corresponding labels (Mark~1, Mark~2, etc.) indicate various VLBI data recording systems. VSOP and RadioAstron markers are added by the author to the plot kindly provided by Alan Whitney (Haystack Observatory), 2015.}
   \label{V_data-rate}   
\end{figure}

\subsection{SVLBI data streaming from an orbital telescope to a correlator}
\label{data_rate}

Continuum observations constitute a sizeable if not a leading part of Space VLBI science case. Given a difficulty in matching the size of a space-borne antenna to that of Earth-based telescopes (sub-section \ref{SVLBI-ant}), and principle limitation of gaining higher sensitivity by decreasing the receiver's system temperature, the only remaining viable resource for increasing the baseline sensitivity in SVLBI is an increase of the recording bandwidth and corresponding digital data rate. In fact, this is the case for Earth-based VLBI too. For that reason, the data rate in VLBI is increasing steadily over the entire history of VLBI from hundreds of kilobits per second in the end of the 1960s to tens of gigabits per second nowadays (Fig. \ref{V_data-rate}). However, both first generation SVLBI missions, VSOP and RadioAstron, were limited in VLBI data rate by 128 Mbps, considerably lower than the maximum available data rate of contemporary Earth-based VLBI systems, especially in the case of RadioAstron operating in 2011--2019. The reason of this apparent deficiency of SVLBI missions is two-fold. 

First, the on-board instrumentation has to be ``frozen'' many years before launch. For both VSOP and RadioAstron, the specification of the data downlink systems has been formulated at about 1990, when the aggregate data rate of 128 Mbps was near the state-of-the-art for Earth-based systems too. It was considered to be too risky to design a space-borne radio telescope with a much higher data rate without a ground-proof in Earth-based VLBI. 

The second reason was more serious. VLBI data of a space-borne telescope cannot be recorded and stored on board the spacecraft due to a too large volume of data and inability to deliver this volume to the ground other than via wide-band downlink. The data must be downloaded in real time. Getting a stream of 128 Mbps from an orbit tens of thousands km above Earth was a challenging task in the 1990s. Even today, almost 30 years later, this is still a non-trivial task requiring dedicated Space VLBI data acquisition stations distributed over the entire range of longitudes as uniformly as possible. There were five dedicated stations supporting VSOP: one JAXA station in Usuda, Japan; three NASA stations in Robledo (Spain), Goldstone (CA, USA) and Tidbinbilla (Australia); and one NRAO station in Green Bank (WVA, USA). RadioAstron operates with two data acquisition stations: Pushchino (Moscow region, Russia) and Green Bank (WVA, USA). In order to accommodate the data stream of 128 Mbps, an unusual for space science applications radio band at about 15 GHz, otherwise designated for remote sensing applications, has been chosen.  Finding a wide enough data downlink band in the strictly regulated electromagnetic spectrum was a problem at the dawn of the SVLBI era. It remains to be a serious technological challenge for a foreseeable future.

All in all, SVLBI data downlink and logistics proved to be a challenging task, somehow resolved in the first generation missions. But for any further steps in SVLBI, whether it employs Space-Earth baselines or Space-Space ones, the complexity of data management should not be underestimated. It is this system that de facto limited the sensitivity of both VSOP and RadioAstron in continuum observations. 

\subsection{The first generation SVLBI major specifications and the way forward}

The first generation SVLBI mission were pushing the envelope of relevant contemporary technologies. Not surprisingly, due to the global character of SVLBI, their major parameters were rather similar (Table \ref{SVLBI-specs}). That said, originally there was an important difference: the VSOP project was conceived as an engineering mission Muses-B aimed at ``live'' verification of technical solutions of large space-borne deployable antennas and wide-band radio links \citep{Nishimura91,Hirosawa91}. Besides, it served as a payload for the first (de facto -- test) flight of the Japanese launcher M-V. The RadioAstron project was adopted for implementation as a science mission right from the beginning \citep{VVA+NSK+SVP79,VVA+NSK81,RZS84}. However, in the first half of the 1990s, both projects have been ``synchronised'' as much as possible. At some point, it was expected that both missions would operate simultaneously. This resulted in standardising the SVLBI data format and wide-band downlink communication radio system at Ku-band (near 15 GHz). Moreover, all five Earth-based VSOP data acquisition stations were designed to be compatible with the RadioAstron radio link system and were expected to work within the RadioAstron mission too.  All that explains obvious similarity of the major parameters of the two implemented missions. Nevertheless, RadioAstron, launched 14 years later than VSOP, featured a wider set of observing bands, the dual-polarisation capability, and a choice of different heterodyning schemes. 
  
\begin{table}
\caption{Major specifications of the two first-generation  SVLBI missions, VSOP \citep{Hir98} (column 2) and RadioAstron \citep{NSK+13} (column 3), and an example of a prospective SVLBI mission ARISE \citet{UlvGuLi97,Ulve99} (column 4).} 
\smallskip
\smallskip
\begin{tabular}{llll}
\hline
Mission                    & VSOP    & RadioAstron & ARISE \\
\hline
Aperture diameter [m]  & 8         & 10              & 25    \\
Observing bands [GHz]  &  1.6, 5    & 0.327, 1.6, 5, 22 & 8, 22, 43, 87 \\
On-board heterodyning & PLL & H-maser, Rb and PLL & PLL \\
Polarisation             & LCP       & LCP \& RCP  & LCP \& RCP    \\
Data rate [Gbps]     & 0.128     & 0.128             & 8     \\
Apogee height [km] & 22,000   & 350,000        & 40,000 \\
Phase-referencing  & No          & No                 & Yes      \\
Highest resolution [$\mu$as] & 300 & 7            & 25   \\                  
\hline
\end{tabular}
\label{SVLBI-specs}
\end{table}

The right column of Table \ref{SVLBI-specs} contains the parameters of the ARISE design study as an illustration of the vector of SVLBI developments as seen around the turn of the century. This study was one of several most advanced attempts to get off the ground a second-generation SVLBI mission. ARISE aimed at increasing SVLBI baseline sensitivity by a factor of $\sim 100$ using a $\sim 2.5$ times larger primary reflector, 64 times higher aggregate data rate and additional enhancement by using phase-referencing technique (effectively -- increasing the integration time). The latter is a very valuable enhancement of the science efficiency of the mission: not only it would improve the detectability of weak ``fringes'', but would also make possible astrometric observations -- something highly desirable but hardly possible with the first generation SVLBI \citep{PoRiMaHi+00}. However, this enhancement does not come for free as it results in a significant complication and increased load on the spacecraft attitude control system. 

It is worth noticing that the drive toward higher angular resolution in the ARISE second generation SVLBI mission was to be addressed by a shift toward higher frequencies (up to 87~GHz) but with a moderate orbit apogee of about 40,000~km, only twice longer than that of VSOP and an order of magnitude smaller than that of RadioAstron. This choice was driven by a desire to get a $(u,v)$-coverage suitable for high-fidelity imaging in observations lasting $~24$ hours or less.

The most challenging parameter of the exemplary second generation mission was the aggregate data rate, 8~Gbps. At the time of this writing, twenty years after the ARISE pre-design study, the sustainable data rate in Earth-based VLBI operations is still four times lower. The Event Horizon Telescope (EHT) is operating with the aggregate data rate of 64~Gbps (EHT collaboration, private communication, 2019). But this is obviously an Earth-based system without real-time transmission of data at that speed. Hopefully, by the time of the implementation of the second generation SVLBI mission, the choice of the data rate will be made in accordance with the contemporary Earth-based bench-mark and accounting for the ``freezing'' delay described above in subsection \ref{data_rate}.

\section{Conclusions and forward look}

The first generation SVLBI missions proved the concept of ultra-high angular resolution with baselines longer than the Earth diameter. Further steps in developments of space-borne radio astronomy, and SVLBI in particular, are inevitable. These developments will have to address several shortcomings of the first generation SVLBI. In cm--dm spectrum domain, the first and foremost is the need to improve the baseline sensitivity. The goal must be to make baselines to a space-borne telescope as sensitive as those for typical Earth-based VLBI systems -- if not better. This is to be the case for both Space-Earth as well as Space-Space baselines. There are two ways of achieving this goal: increase of the telescope diameters from the already demonstrated values of $\sim 10$~m and more than 10-fold increase of the VLBI data rate. Both remedies can work for continuum observations, but the latter one has only limited applicability for spectral line observations (only via more sensitive phase-referencing to a continuum near-by calibrator source).  There is also a limitation to the applicability of the increased data rate to VLBI observations at frequencies lower than $\sim 1$~GHz due to the limitation of the available bandwidth.

Functional operational constraints of globally distributed SVLBI systems should not be underestimated. For both VSOP and RadioAstron there were attempts to evaluate and minimise the impact on the mission operations such the factors as space-borne attitude limitations, accessibility of Earth-based data acquisition stations, availability of co-observing Earth-based radio telescopes. These efforts were only partially successful: for both missions, but especially for RadioAstron, the actual functional constraints (e.g., thermal conditions of the on-board radio link devices) made significant impact beyond expectations thus limiting the overall time-efficiency of science. This is, perhaps, a trivial conclusion, yet very relevant for such the diverse and complicated tool as SVLBI: a system approach for its design should be exercised right from the beginning of its development. 

Over the past nearly forty years, the main attention in developing SVLBI concepts was given to ``traditional'' for Earth-based VLBI frequency bands from 0.3 to 43 GHz. Besides mentioned above enhancements in sensitivity, a truly new quality for these meter to ``several millimeter" wavelengths SVLBI systems could be achieved by multi-spacecraft systems employing Space-Space baselines. The advantage of such the systems have been at the center of attention in such the studies as iARISE \citep{Murphy+05}, the Chinese SVLBI project studies at millimeter \citep{Hong+13} and longer wavelength \citep{AnTao+18,AnTao+19} domains. Such the SVLBI systems operating without Earth-based telescopes might offer three new attractive features: (i) they do not suffer the impact of atmosphere or ionosphere scattering; (ii) in principle, depending on the orientation of space-borne telescopes' orbits, they can achieve longer baselines and faster $(u,v)$-coverage, and (iii) they are free of Earth horizon visibility limitations of Earth-based telescopes. 

Space VLBI tends to broaden the range of operational frequencies in both directions, to millimeter and decameter (and longer) wavelengths.

Earth-based VLBI observations at millimeter wavelengths are becoming increasingly difficult as frequencies reach 300~GHz and higher values. At present, the Global mm-VLBI Array (GMVA)\footnote{https://www3.mpifr-bonn.mpg.de/div/vlbi/globalmm/ (accessed 15.02.2019)} and Event Horizon Telescope (EHT)\footnote{https://eventhorizontelescope.org (accessed 15.02.2019)} operate at short millimeter wavelengths and attempt to venture into sub-millimeter domain. The main difficulties of these activities are scattering and absorption in the Earth atmosphere. Space-borne interferometers are free from these problems. Several pre-design studies addressed mm-SVLBI issues over the past two decades: ARISE \citep{Ulve00}, VSOP-2 \citep{Hagi+09} (Part~2) and Millimetron \citep{NSK+14}. The current Special Issue presents the current status of Millimetron \citep{Andria+19} and several new concepts \citep{Fish+19,Linz+19,Kudria+19}. All these concepts face many engineering challenges. But overall progress of space technology and other relevant developments in science instrumentation will steadily move them toward acceptable for implementation Technology Readiness Level (TRL) grades.

At the opposite end of the electromagnetic spectrum, long and ultra-long wavelengths (ULW), the major enemy of Earth-based radio astronomy (and radio interferometry as well) is the Earth ionosphere, which is practically opaque at frequencies lower than $\sim15$~MHz (wavelengths longer than $\sim20$~m). For that reason, the ULW spectral domain is the last unexplored region of cosmic electromagnetic emission. In order to open up this last hitherto closed window into the electromagnetic Universe, the telescopes must be placed above ionosphere. In the recent years several attempts to create ULW facilities intensified. An important role in these studies play the Moon as a natural shield from anthropogenic radio frequency interference (RFI). It is expected that the first demonstration of ULW VLBI will be attempted in 2019 in the framework of the Chinese--Dutch experiment NCLE aboard the Chinese Lunar mission Chang'E-4 \citep{Jia+18}. Several other ULW VLBI initiatives and projects are under development with the aim of becoming operational in the coming decade  (\citet{Boonstra+16,Belov+18,Bentum+19} and references therein).

\section{Acknowledgements}

All three implemented to date Space VLBI systems were massive undertakings involving hundreds of engineers and scientists. But in each case, several leaders played the decisive role: Gerry Levy (TDRSS Orbital VLBI demonstration), Hisashi Hirabayashi and Haruto Hirosawa (VSOP--HALCA), Nikolai Kardashev and Vladimir Andreyanov (RadioAstron). The author is grateful to many colleagues involved in the VSOP and RadioAstron missions with whom he has been privileged to work for several decades. The TDRSS OVLBI experiment was led by the NASA JPL with key contribution by the TDRSS organisation. The VSOP Project was led by the Institute of Space and Astronautical Science of the Japan Aerospace Exploration Agency, in cooperation with many organisations and radio telescopes around the world. The RadioAstron project is led by the Astro Space Center of the Lebedev Physical Institute of the Russian Academy of Sciences and the Lavochkin Scientific and Production Association under a contract with the State Space Corporation Roscosmos, in collaboration with partner organisations in Russia and other countries. The author is grateful to S\'{a}ndor Frey for careful reading of the manuscript and anonymous referees for useful comments and suggestions.


\end{document}